\documentclass[structabstract]{aa}
\usepackage{rotating}
\usepackage{graphicx}      
\usepackage{txfonts}   
\input epsf.sty   

\begin{document} 
\title{Absolute emission altitude of pulsars:  PSRs B1839+09, 
  B1916+14 and B2111+46} 
 \author{R. M. C. Thomas 
          \inst{1}
          \and
 R. T. Gangadhara
\inst{2}
          }
\institute{ National Center for Radio Astrophysics, Pune -- 411007, India \\
\email{mathew@ncra.tifr.res.in} 
 \and
Indian Institute of Astrophysics, Bangalore -- 560034, India\\ 
\email{ganga@iiap.res.in } }

\abstract {} { We study the mean profiles of the multi--component
  pulsars PSRs~B1839+09, B1916+14 and B2111+46.  We estimate the
  emission height of the core components, and hence find the absolute
  emission altitudes corresponding to the conal components. } { By
  fitting Gaussians to the emission components, we determine the phase
  location of the component peaks.  Our findings indicate that the
  emission beams of these pulsars have the nested core--cone
  structures.  Based on the phase location of the component
  peaks, we estimate the aberration--retardation (A/R) phase shifts in
  the profiles.  Due to the A/R phase shift, the peak of the core
  component in the intensity profile and the inflection point of the
  polarization angle swing are found to be symmetrically shifted in
  the opposite directions with respect to the meridional plane in such
  a way that the core shifts towards the leading side and
  the polarization angle inflection point towards the trailing side. }
 { We have been able to
  locate the phase location of the meridional plane and to estimate the
  absolute emission altitude of both the core and the conal components
  relative to the neutron star center, using the exact expression for the
  A/R phase shift given by Gangadhara (2005).}  {}
  
\authorrunning{Thomas \and  Gangadhara}
\titlerunning{Absolute Emission Altitude of Pulsars}
 \maketitle

\section{Introduction}  
Pulsar radio emission is understood to be emitted by the relativistic
plasma accelerated along the dipolar magnetic field lines (e.g.,
Ruderman \& Sutherland 1975).  Among the various models proposed for
pulsar emission, the coherent curvature radiation has turned out to be
an effective mechanism for explaining some of the important pulsar
radiation properties.  The common occurrence of an odd number of
components in the mean pulsar profiles has lead to the nomenclature of
a nested conal structure for the pulsar emission beam (e.g., Rankin
1983a; Rankin 1993).  However, Lyne \& Manchester (1988) suggested
that the emission within the beam is patchy, i.e., the distribution of
component locations within the beam is random rather than organized in
one or more hollow cones. Also studies by Mitra \& Deshpande (1999)
indicate that the structure of the pulsar emission beam is more likely
to be nested hollow cones.  Gangadhara \& Gupta (2001, hereafter
GG01), and Gupta \& Gangadhara (2003, hereafter GG03) showed that the
prevalent picture of emission cones axially located around the central
core component is a suitable model for explaining the core-cone
structure of the pulsar emission beam.

A long--standing question in pulsar astronomy has been the location of
the radio emission region in the magnetosphere. In the literature,
there are mainly two types of methods proposed for estimating the
radio emission altitudes: (1) {\it a purely geometric method,} which
assumes the pulse edge is emitted from the last open field lines
(e.g., Cordes 1978; Gil \& Kijak 1993; Kijak \& Gil 2003), (2)~{\it a
  relativistic phase shift method,} which assumes that the asymmetry
in the conal components phase location relative to the core is due to
the aberration-retardation phase shift (e.g., GG01, Gangadhara 2005,
hereafter G05). Both methods have merits and demerits: the first
method has an ambiguity in identifying the last open field lines,
while the latter is restricted to the profiles in which the core-cone
structure can be clearly identified.  The emission heights of PSR
B0329+54 given in GG01, six other pulsars in GG03 and the revised ones
by Dyks, Rudak \& Harding (2004, hereafter DRH04) are all relative to
the emission height of the core, which is assumed to be zero. However,
the core emission is believed to originate from lower altitudes than
that of the conal components (e.g., Blaskiewicz et~al. 1991; Rankin
1993).  Hoensbroech \& Xilouris (1997) estimated the emission heights
at high frequency radio profiles for a set of pulsars.  They suggested
that the emission heights at high frequency can set an upper limit for
the core emission height.

By assuming a fixed emission altitude across the pulse, Blaskiewicz
et~al.~(1991, hereafter BCW91) presented a relativistic rotating
vector model. The results of this purely geometric method are found to
be in rough agreement with those of BCW91.  However, the relativistic
phase shift method clearly indicates that the emission altitude across
the pulse window is not constant (GG01; GG03; DRH04; Johnston \&
Weisberg 2006; Krzeszowski et al. 2009).

By considering the relativistically beamed radio emission in the
direction of the magnetic field line tangents, Gangadhara (2004,
hereafter G04) solved the viewing geometry in an inclined and slowly
rotating dipole magnetic field. A more exact expression for the
relativistic phase shift is given in (G05), which also includes the
phase shift due to polar cap currents. In the present work, we analyze
the mean profiles of PSRs B1839+09 and B1916+14 at 1418 MHz, and
PSR~B2111+46 at 610 MHz and 1408 MHz, to estimate the absolute
emission height of the pulse components. In Sect.~2, we give a method for
estimating the absolute emission height of pulse components.
   
\section{Method for estimating the absolute emission heights}
    \label{sec:method}
The work of BCW91 generalized the rotating vector model (RVM) to
include the relativistic effects due to rotation.  According to their
model, the centroid of the intensity profile advances to an earlier
phase by $\sim r/r_{\rm LC},$ while the polarization position angle
inflection point (PPAIP) is delayed to a later phase by $\sim 3\,
r/r_{\rm LC},$ where $r$ is the radial distance from the center of
neutron star and $r_{\rm LC}$ is the light cylinder radius.  After
estimating the width of the pulse at $\approx 10\%$ intensity level
and by fitting the relativistic rotating vector model, they estimated
the phase shift between the centroid of the profile and the PPAIP. But
the retardation phase shift was ignored in BCW91, as they assumed a
constant emission height across the pulse.  In GG01, GG03, Johnston \&
Weisberg (2006) and Krzeszowski et al. (2009) it was shown though that
the emission altitude is not constant across the pulse, and hence
retardation has to be taken into account while estimating the A/R
phase shifts.  Further, Dyks et~al. (2004) showed that the centroid of
the intensity profile advances by $\sim 2\,r/r_{\rm LC}$ while the
PPAIP is delayed by $\sim 2\,r/r_{\rm LC}$ due to A/R effects with
respect to the meridional plane.
\begin{figure}
\begin{center}
\epsfxsize= 8 cm
\rotatebox{0}{\epsfbox{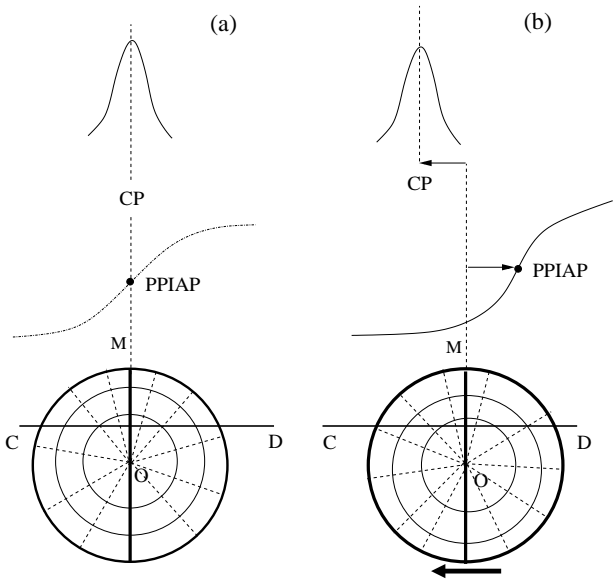}}
\caption[short_title]{\small Schematic diagram showing the A/R phase
  shift between the core peak (CP) and the polarization position angle
  inflection point (PPAIP).  The panel (a) for the co-rotating frame,
  where the phases of PPAIP and CP coincide with that of the
  meridional plane (M), and (b) in laboratory frame, due to A/R
  effects, both the PPAIP and the CP are symmetrically shifted in the
  opposite directions with respect to M.
\label{PPAIP}}     
\end{center}             
\end{figure}    

By solving the viewing geometry in the dipole magnetic field,
Gangadhara (2005) showed that instead of a centroid of intensity
profile, the phase shift of the central (core) peak (CP) relative to
the meridional plane must be considered for estimating the A/R phase
shifts. In the co-rotating (non-rotating) case both core and PPAIP
originate from the same phase (meridional plane M, see
Fig. 1a). Whereas in the observer (laboratory) frame, the CP shifts to
the earlier phase and the PPAIP to the later phase by the same
magnitude (see Fig.~1b).  Hence to find the absolute emission height
of the profile components including the central (core) component, we
adapted the method of G05 to {\it consider the CP instead of the
  centroid of pulse (BCW91) for estimating the A/R phase shift.}  It
is logical to presume that the aforesaid $r$ should be the same for
the origin of the central (core) component and the PPAIP.  Or stated
otherwise, the phase difference $\Delta\phi'_{\rm CP}=\phi'_{\rm
  core}-\phi'_{\rm PPAIP}\sim 0,$ in the co-rotating frame, where $
\phi'_{\rm core}$ is the phase location of the core peak while
$\phi'_{\rm PPAIP}$ is the phase location of the PPAIP. 

\begin{figure}
\begin{center}
\epsfxsize= 9 cm
\rotatebox{0}{\epsfbox{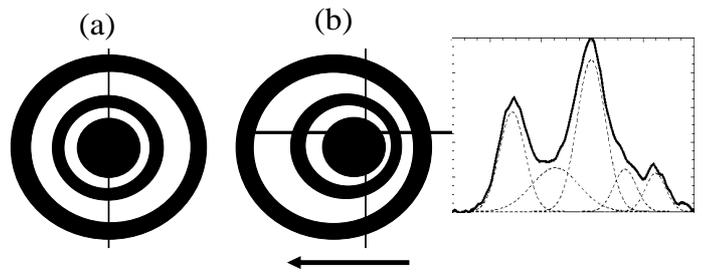}}
\caption[short_title]{\small Schematic diagram to show the probable
  distribution of emission patterns across the pulsar beam.  The beam
  cross sections as they appear in (a) the co-rotating frame and (b)
  the laboratory frame, where the cones are not coaxial with the
  central core because of A/R retardation effects.  The thick
  horizontal line represents the direction of the tracing of the
  line-of-sight across the beam. The resultant intensity profile is
  shown in the adjoining box. The vertical line denotes the meridional
  plane, and the thick arrow represents the direction of the pulsar
  rotation.
\label{AB_RET}}
\end{center}
\end{figure}

As illustrated
in the Fig.~\ref{AB_RET}, the panel (a) depicts the cross section of
the emission region in the co-rotating frame, while panel (b) shows for the
same in the observer's frame.  The thick arrow represents the
direction of the rotation, and the thinner line the sweep of the
line-of-sight. The shaded ring-like region represents the nested conal
emission regions and the central circle the core emission region. The
sweep of the line-of-sight across the depicted region causes the core
peak and the PPAIP to be separated by a roughly equal measure ($\sim 2
r/r_{\rm LC}$) in opposite directions from the meridional plane (M),
as illustrated in Fig.~\ref{PPAIP}.  The resultant intensity profile,
which characterizes a sum total of emissions after the line-of-sight
sweeps across the emission region, is shown in the adjoining box on the
right hand side.

\section{Application of the method}
The implementation of the aforesaid method to estimate the absolute
emission height of the core and conal components is described below.  We
considered the mean profiles of PSRs~B1839+09, B1916+14 and B2111+46
for our study, as they exhibit a clearly identifiable core and smooth
polarization-position-angle (PPA) swing.  We obtained the data of PSRs
B1839+09 and B1916+14 from Everette \& Weisberg (2001), and those of
PSR B2111+46 from EPN data base.

\subsection{Longitude of the core peak}
\label{subsec:corepeak}
\begin{figure} 
\begin{center} 
\epsfxsize= 8 cm   
\epsfysize=0cm
\rotatebox{0}{\epsfbox{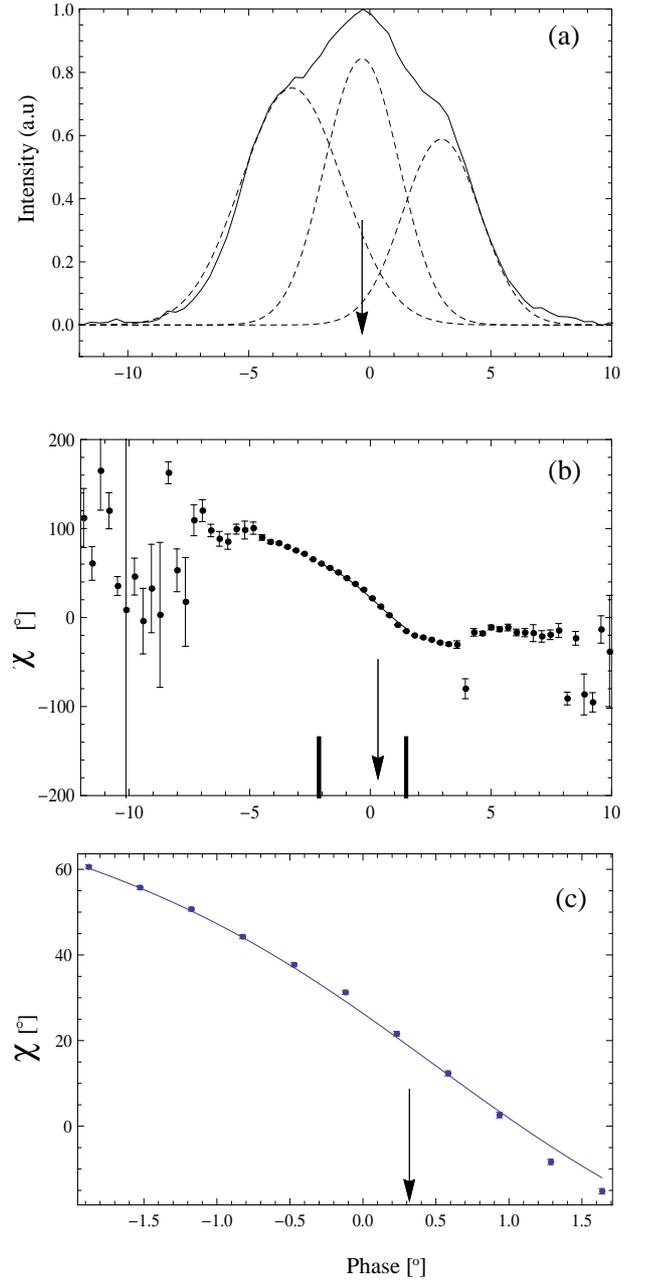}}
\caption[short_title]{\small Intensity profile of PSR B1839+09 at
  1418 MHz, fitted with the Gaussians to the sub-pulse
  components. In panel (a) the continuous line represents the observed
  mean profile while the broken line curves represent the fitted
  gaussians.  The arrow points to the phase of the core peak. In panel
  (b) the corresponding polarization angle ($\chi$) is fitted with
  relativistic RVM curve (BCW91) curve. The arrow points to the phase
  of inflection point (PPAIP), and the vertical lines mark the region
  of phase over which the BCW91 curve is fitted. In panel (c) the zoomed
  out region demarcated by the vertical lines in panel (b), is plotted
  and the arrow points to the PPAIP.
\label{freq_1839}}
\end{center}      
\end{figure}

We fitted Gaussians to the pulse components to resolve the individual
components based on the method developed by Kramer et~al. (1994), and
the profiles are given in Figs.~\ref{freq_1839} --
\ref{freq_1408}. Hence the peak-phase locations of the individual
components are resolved.  The broken line curves in panel (a) indicate
the fitted Gaussians.  The phase location of the core peak $(\phi'_{\rm
  core})$ is marked with an arrow and is tabulated in
Table~\ref{tabcore}. The PPA is plotted in panel (b), and the vertical
lines mark the fitted region of the curve.  The panel (c) shows the
zoomed-out region of the PPA within the region of the fit.  The arrow
points to the PPAIP in both panels (b) and (c).

The location of the central component (core) is expected to appear at M
for an observer in the co-rotating frame as explained in
Sect.~\ref{sec:method} and illustrated in Fig.~1(a).  But for an observer
in the laboratory frame, the core emission will be advanced to an
earlier phase by $\delta\phi'_{\rm core}\sim - 2\, r_{\rm core}/
r_{\rm LC}$ and the corresponding PPAIP delayed to a later phase by
$\delta\phi'_{\rm PPAIP}\sim 2\, r_{\rm core}/r_{\rm LC},$ where
$r_{\rm core}$ is the emission height of the core. Then the A/R phase
shift of the core with respect to M is $\delta\phi'_{\rm core}=\Delta
\phi'/2,$ and the parameters related to core emission are given in
Table~\ref{tabcore}.  The frequency $\nu$ of each data set is given in
col.~2, and the phase shifts $\delta\phi'_{\rm core}$ and
$\delta\phi'_{\rm PPAIP}$ in cols.~3 and 4.
       
\subsection{The relativistic RVM fit} 
 We fitted the relativistic RVM (BCW91) to the region of the PPA data,
 which corresponds to the core emission, the region over which the
 core polarization is significantly higher than that over the
 adjacent conal regions.  We could see from the profiles that the
 pulse phase range falling within the full-width-at-half-maximum
 (FWHM) of the central-fit-Gaussian can clearly bracket out the
 emissions that dominate the core.  The core emission is found to be
 dominant within 10\% intensity levels of the central Guassian of the
 1418 MHz data of PSR B1916+14, and hence we included that region for
 fitting the BCW91 curve for that profile.  Additional reasons for
 restricting the fit region are discussed in detail in later sections.

\subsubsection{Longitude of polarization position angle inflection point}
 We invoked the guess values for the emission height $r$ and the phase
 $\phi'_{\rm PPAIP}$ to fit the BCW91 curve. The polarization angle
 data, falling within the FWHM of the core, are fitted with a 6th
 degree polynomial: $\psi(\phi')=a_{0}+a_{1} \phi' +a_{2} \phi'^2 +
 \cdot\cdot\cdot ,$ where $a_{0},~a_{1},~a_{2}, \cdot\cdot\cdot$ are
 the fit coefficients, and $\phi'$ is the pulse phase in degrees. We
 differentiated the fitted polynomial and found the maximum of $\vert
 d\psi/d\phi'\vert$ that gives the guess value of $\phi'_{\rm PPAIP}.$
 In the next step, the PPA data were fitted with the relativistic RVM
 curve (BCW91) using the following expression
 \begin{equation}\label{eq_{BCW0}}
 \psi_{\rm BCW}=\psi_0+ \tan^{-1}\left[\frac{\sin\alpha \, \sin(\Omega
     t) -3 (r/r_{\rm LC})
     \sin\zeta}{\sin\beta+\sin\alpha\cos\zeta(1-\cos(\Omega
     t))}\right]~,
 \end{equation}                                                                 
 where $\zeta=\alpha+\beta.$ The inclination angle of the magnetic
 axis relative to rotation axis is $\alpha,$ and the impact angle of
 the line-of-sight relative to the magnetic axis is $\beta.$ The
 fitted PPA data are shown in the panel (b) of
 Figs.~\ref{freq_1839}--\ref{freq_1408}.

 The parameter $\psi_0$ is inserted in the above Eq.~(\ref{eq_{BCW0}})
 in order to offset the possible arbitrary and constant `vertical
 shift' that the raw PPA data might have. This is due to the arbitrary
 value of the projection of the rotation axis in the sky plane. Since
 we are interested in finding $r,$ which is expected to be relatively
 constant in the region of fit, we assumed it to be independent of
 $\phi'\,\,(=\Omega \,t)$, and hence it is taken as a fitting
 parameter.  The vertical shift in the raw PPA data, in the first
 step, was brought closer to zero by finding a trial value for the
 PPAIP on the vertical axis ($\psi_{\rm T}$) from the polynomial fit,
 and thereafter the data were shifted vertically so that $\psi_{\rm
   POL}\rightarrow\psi_{\rm POL}-\psi_{\rm T},$ where $\psi_{\rm POL}
 $ is the observed polarization angle.  These PPA data were fitted with
 the Eq.~(\ref{eq_{BCW0}}) keeping $r$ and $\psi_0$ as the free
 parameters; hence allowing two degrees of freedom for the fit
 function, i.e., allowing the fit function to `adjust' in both
 the vertical (through the free parameter $\psi_0$) and horizontal
 direction (through free parameter $r$) for a good fit.  The fit
 procedure was repeated consecutively a few times, with the values of
 $r$ and $\psi_0$ from the preceding fit as guess values for the final
 convergent and stable values.  Thus  $\phi'_{\rm PPAIP} $ was found
 from the fit, and the corresponding values are shown in
 Table~\ref{tabcore}.  The geometric angles $\alpha$ and $\beta$ were
 not invoked as fit parameters in Eq.~(\ref{eq_{BCW0}}) because we 
 used their predetermined (published) values in the BCW91 fit
 function.  We used the values of $\alpha$ and $\beta$ given by
 Everette \& Weisberg (2001) for PSRs~B1839+09 and B1916+14, and for
 PSR~B2111+46 by Mitra \& Li (2004, hereafter ML04).

\subsection{The longitude of the meridional plane}\label{subsec:Mplane}

The CP and PPAIP of the polarization angle appear at M in the
co-rotating frame, as indicated by Fig.~\ref{PPAIP}(a). But they are
symmetrically shifted in opposite directions from M due to the A/R
phase shift for an observer in the laboratory frame, as indicated by
Fig.~\ref{PPAIP}(b). Note that the phase location of M is invariant
with respect to the rotation effects, or in other words, both the CP and
the PPAIP come closer to M at smaller $r,$ and move away from it at
larger $r.$ If $\phi'_{\rm core}$ and $\phi'_{\rm PPAIP}$ are the
estimated phases of the CP and the PPAIP, then their phase difference
($\Delta\phi'$) is given by $\Delta\phi' =\phi'_{\rm PPAIP}-\phi'_{\rm
  core},$ and the meridional plane M is situated at the mid point  
between CP and PPAIP.  Hence the phase of M is given by $\phi'_{\rm      
  M}=\phi'_{\rm core}+(\Delta\phi'/2).$ In Figs.~\ref{freq_1839} --
\ref{freq_1408} the phases are shifted by $\phi'_{\rm M},$ so that M
appears at the zero phase and an arrow points to the phase location
$\phi'_{\rm PPAIP}$ in the polarization angle panels (b).
 
\begin{table*}   
\caption{{Core emission geometry parameters of PSRs B1839+09, B1916+14
    and B2111+46 }
\label{tabcore}}
\begin{tabular}{crrrrrrrrrrrrrrrr} 
\hline
\hline
\\
\multicolumn{1}{c}{Pulsar}& \multicolumn{1}{c}{$\nu$}  
&\multicolumn{1}{c}{$\delta\phi'_{\rm core}$}
&\multicolumn{1}{c}{$\delta\phi'_{\rm PPAIP}$}
 & \multicolumn{1}{c} {$r_{\rm core}$ }
& \multicolumn{1}{c}{$r_{\rm core}$}
 & \multicolumn{1}{c}{$\chi^2$}
 & \multicolumn{1}{c}{SR}
& \multicolumn{1}{c}{$s/s_{\rm lof}$}    \\
\multicolumn{1}{c}{PSR B}& 
\multicolumn{1}{c}{(MHz)} & \multicolumn{1}{c}{($^\circ$)}  &\multicolumn{1}{c}{ ($^\circ$)}
  &\multicolumn{1}{c}{(km)$^a$}
 & \multicolumn{1}{c}{$(\%\,\, r_{\rm LC})$}     
& &
 (\%)$^b$
&\multicolumn{1}{c}{foot value}  
  \\
 \multicolumn{1}{c}{(1)} & \multicolumn{1}{c}{(2)}  &\multicolumn{1}{c}{ (3)}
 &\multicolumn{1}{c}{(4)}  &
  \multicolumn{1}{c}{(5)} & \multicolumn{1}{c}{(6)}&  \multicolumn{1}{c}{(7)}&
  \multicolumn{1}{c}{(8)} & \multicolumn{1}{c}{(9)}
  \\
\hline
  \\ 
1839+09   &  1418 &     -0.32$\pm$0.07   &   0.32$\pm$0.01   &   50$\pm$5   & 
            0.28$\pm$0.03   &   1.02   &   100   &   0.51$\pm$0.03\\ \\
1916+14   &  1418 &     -0.23$\pm$0.02   &   0.23$\pm$0.01   &   95$\pm$4   &   0.20$\pm$0.01  
          &   28.25   &   94   &   0.26$\pm$0.01 \\ \\
2111+46   &  610 &     -1.33$\pm$0.02   &   1.33$\pm$0.20   &   503$\pm$38   &   1.04$\pm$0.08   
          &  5.7   &   100   &   0.13$\pm$0.00\\ \\
          &  1408 &     -0.23$\pm$0.08   &   0.23$\pm$0.29   &   83$\pm$54   &   0.17$\pm$0.11   
          &   2.05   &  95  &   0.51$\pm$0.17\\ \\
\hline
\end{tabular} 
  \\ \\ \tiny{$^a$Emission heights computed using the exact formula
    (G05).}  \\ \tiny{$^b$The percentage of the standardized residuals
    (SR) with value of $-2$ and $+2$.}
\end{table*}

\begin{figure}
\begin{center}
\epsfxsize= 8 cm 
\epsfysize=0cm   
\rotatebox{0}{\epsfbox{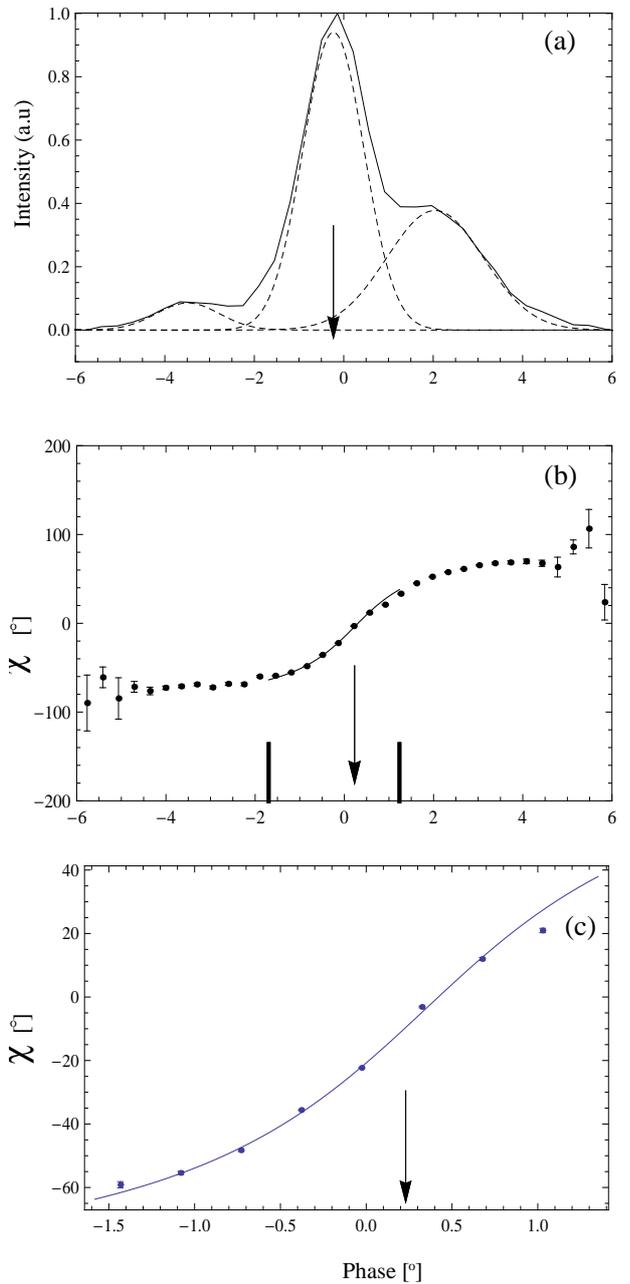}} 
\caption[short_title]{\small Intensity profile of PSR B1916+14 at
  1418 MHz. See the caption of Fig.~\ref{freq_1839} for details.
\label{freq_1408}} 
\end{center} 
\end{figure}   

\subsection{The A/R phase shift of the core }
The location of the central component (core) should appear at M for an
observer in the co-rotating frame as explained in
Sect.\ref{sec:method}.  But for an observer in the laboratory frame,
the core emission will be advanced to an earlier phase by
$\delta\phi'_{\rm core}\sim - 2\, r_{\rm core}/ r_{\rm LC}$ and the
corresponding PPAIP be delayed to a later phase by $\delta\phi'_{\rm
  PPAIP}\sim 2\, r_{\rm core}/r_{\rm LC},$ where $r_{\rm core}$ is the
emission height of the core. Then the A/R phase shift of the core with
respect to M is $\delta\phi'_{\rm core}=\Delta \phi'/2.$
\subsection{Phase locations of the core and cone component peaks }
\label{subsec:comppeaks}

 We found the peak locations of the individual components by fitting
 Gaussians to the mean profiles of the three pulsars.  As mentioned in
 Sect.~\ref{subsec:Mplane}, the data were shifted by $\phi'_{\rm M}$
 so that M appears at zero phase. Because the A/R effects are absent
 in the co-rotating frame, the conal components are expected to be
 symmetrically located on either sides of meridional plane as
 indicated by Fig.~\ref{AB_RET}(a). But in the laboratory frame, the
 cones are advanced to an earlier phase due to the A/R effects and are
 hence asymmetric with respect to the core location as indicated by
 Fig.~\ref{AB_RET}(b).  The meridional plane M is taken to be at the
 zero phase, and the measured phases are, therefore, the absolute
 phases with respect to M.  Accordingly the estimated emission heights
 are the absolute emission altitudes with respect to the center of the
 neutron star. 

 Let $\phi'_{\rm L}$ and $\phi'_{\rm T}$ be the peak locations of the
 conal components on the leading and trailing sides of a pulse profile,
 respectively. Then, using the following equations, we estimate the
 A/R phase shift $(\delta\phi')$ of the cone center with respect to M,
 and the phase location $(\phi')$ of the component peaks in the
 absence of the A/R phase shift, i.e., in the co-rotating frame (see
 Eq.~(11) in GG01):
\begin{equation}\label{dpphip}
\delta\phi'=\frac{1}{2}(\phi'_{\rm T}+\phi'_{\rm L})~,\quad\quad
\phi'=\frac{1}{2}(\phi'_{\rm T}-\phi'_{\rm L})~.
\end{equation} 
 
\section{The absolute emission heights}    
\subsection{Emission height of the core}
The core emission height was computed by using the $\delta\phi'_{\rm
  core}$ in the expression for the A/R phase shift given by G05 (see
Eq.~(45)).  The parameters related to the core emission are given in
Table~\ref{tabcore}. In Col.~6 the emission heights are given as a
percentage of the light cylinder radius $r_{\rm LC}$. It shows that the
core emission in the radio band occurs over a range of altitude
spanning from 0.2 to 1 per cent of  the light cylinder radius. The radio
frequency $\nu$ of each data set is given in Col.~2, and the phase
shifts $\delta\phi'_{\rm core}$ and $\delta\phi'_{\rm PPAIP}$ in
Cols.~3 and 4, respectively.  The values of $\chi^2$ and the standard
residuals obtained are given in Cols.~7 and 8, respectively.  The
foot location of magnetic field lines on the polar cap relative the
magnetic axis are given in Col.~9.
\begin{figure}
\begin{center}
\epsfxsize= 8 cm
\rotatebox{0}{\epsfbox{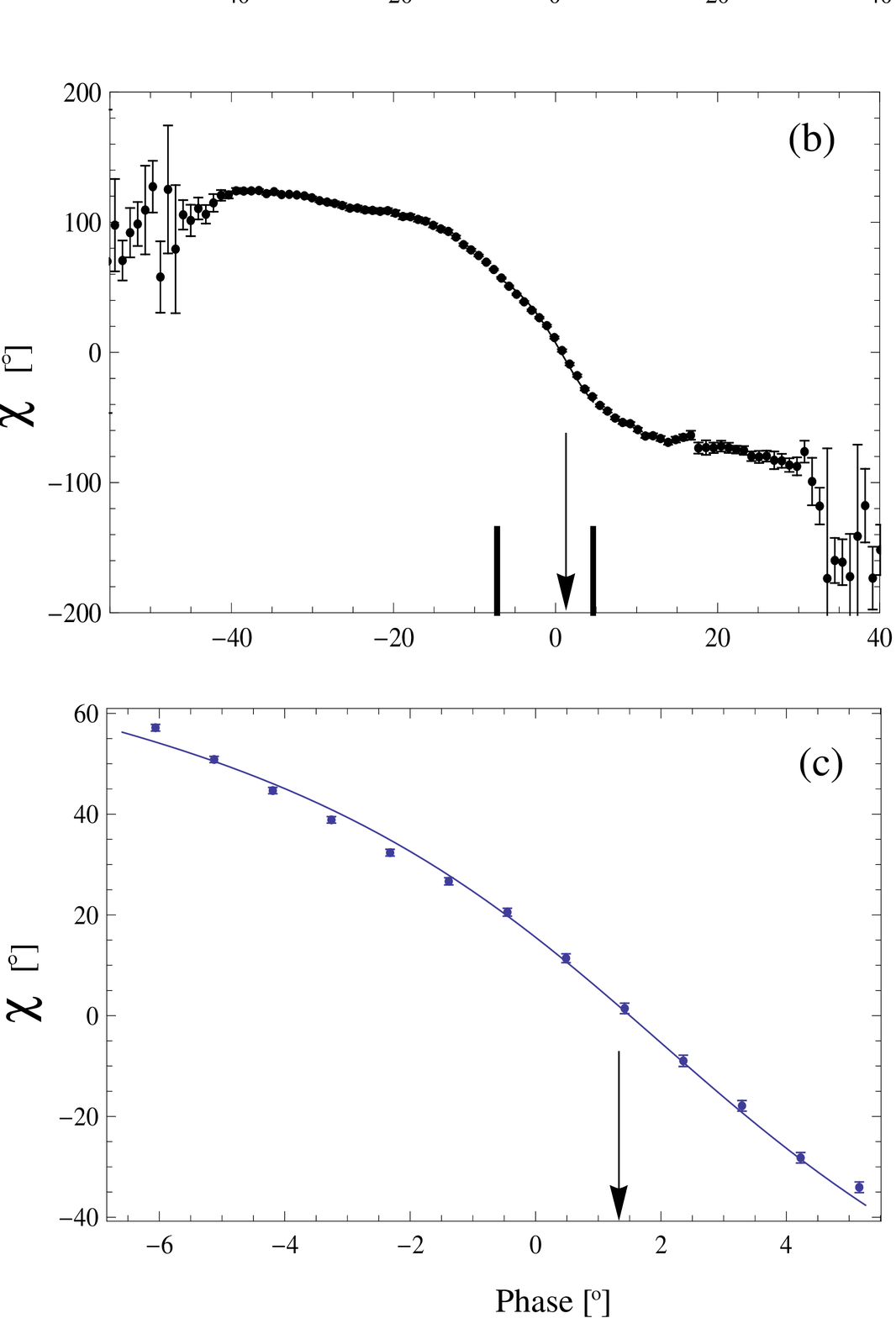}}
\caption[short_title]{\small
Intensity profile  of PSR B2111+46 at 610 MHz. See the caption of 
Fig.~\ref{freq_1839} for details.
\label{freq610}}
\end{center}
\end{figure}

\begin{figure}
\begin{center}
\epsfxsize= 8 cm
\rotatebox{0}{\epsfbox{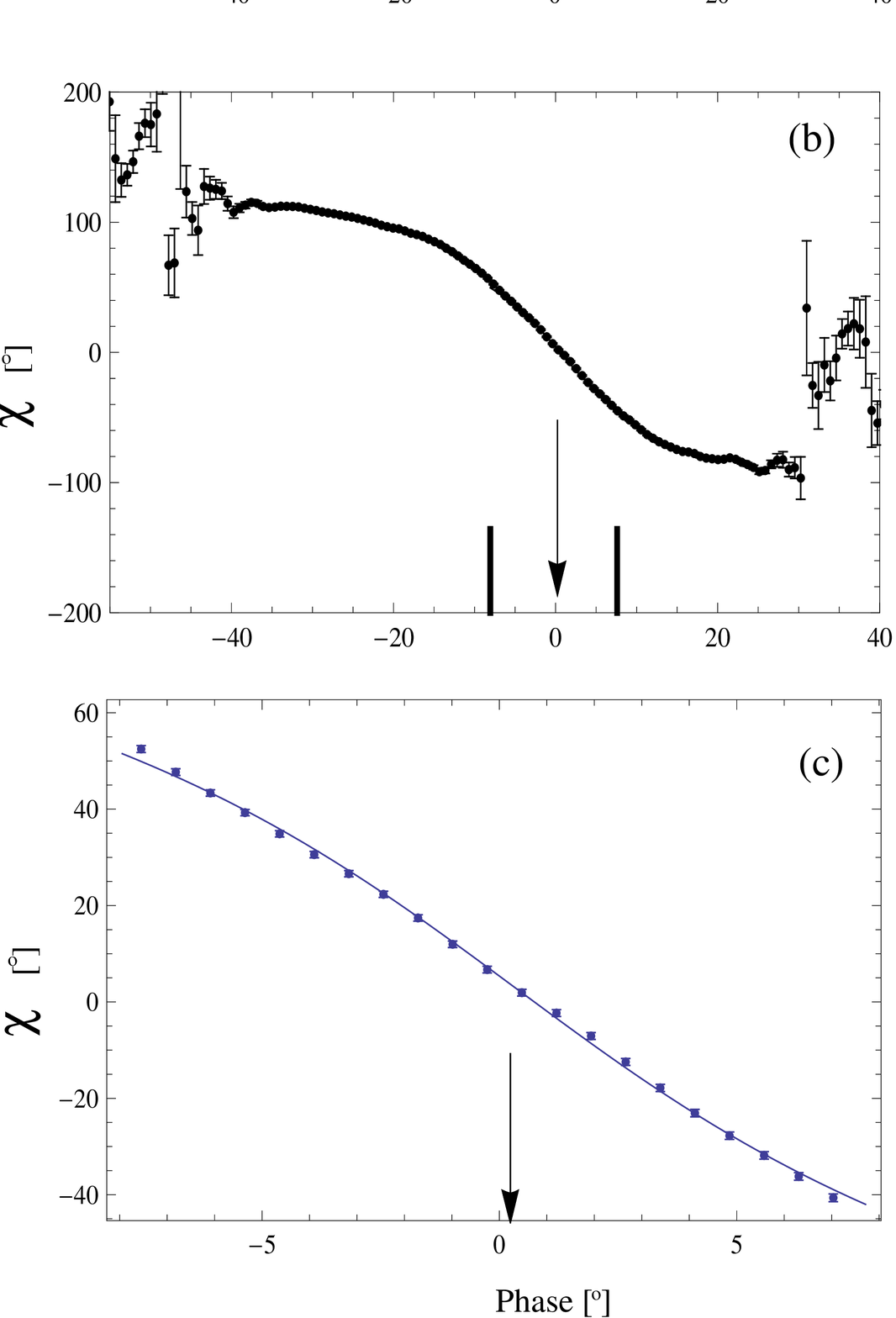}}
\caption[short_title]{\small
Intensity profile  of PSR B2111+46 at 1408 MHz. See the caption of 
Fig.~\ref{freq_1839} for details.
\label{freq_1408}}
\end{center}
\end{figure}

\begin{table*}
\caption{Conal emission geometry parameters of PSRs B1839+09,
B1916+14 and B2111+46 }
\label{tabcomp}
\begin{tabular}{ccrrrrrrccrrr}
\hline \hline 
\\ 
\multicolumn{1}{c}{Pulsar} & \multicolumn{1}{c}{$\nu$} &
\multicolumn{1}{c}{Cone} &\multicolumn{1}{c}{$\phi'_{\rm L}$} &
\multicolumn{1}{c}{$\phi'_{\rm T}$} &
\multicolumn{1}{c}{$\delta\phi'$} &\multicolumn{1}{c} {$\Gamma$}
&\multicolumn{1}{c}{$r$} &\multicolumn{1}{c}{$r$} &
\multicolumn{1}{c}{$s/s_{\rm lof}$} \\ \multicolumn{1}{c}{PSR B} &
\multicolumn{1}{c}{(MHz)} & \multicolumn{1}{c}{(No.)}  &
\multicolumn{1}{c}{($^\circ$)} & \multicolumn{1}{c}{($^\circ$)} &
\multicolumn{1}{c} {($^\circ$)} &\multicolumn{1}{c} {($^\circ$)}  & \multicolumn{1}{c}{(km)$^a$}
 &\multicolumn{1}{c}{$(\%\,\, r_{\rm LC}$)}  &
\multicolumn{1}{c}{foot value} \\ \multicolumn{1}{c}{(1)} &
\multicolumn{1}{c}{(2)} & \multicolumn{1}{c}{(3)} &
\multicolumn{1}{c}{(4)} & \multicolumn{1}{c}{(5)} &
\multicolumn{1}{c}{(6)} & \multicolumn{1}{c}{(7)} & 
\multicolumn{1}{c}{(8)} & \multicolumn{1}{c}{(9)} &
\multicolumn{1}{c}{(10)} \\ \hline \\
 1839+09   &  1418&1    &    -3.75$\pm$0.15    &    2.96$\pm$0.18   &    -0.39$\pm$0.12   &  
              4.06$\pm$2.30   &    63$\pm$19   &    0.34$\pm$0.10   &    0.80$\pm$0.10\\ \\
 1916+14   &  1418&1    &    -3.49$\pm$0.05    &    2.04$\pm$0.07   &    -0.72$\pm$0.04   &  
              2.63$\pm$1.00   &    299$\pm$18   &    0.63$\pm$0.04   &    0.39$\pm$0.01\\ \\
 2111+46  &  610&1    &    -14.78$\pm$0.45    &    11.79$\pm$0.31   &    -1.49$\pm$0.27   &  
             2.85$\pm$1.10 &    581$\pm$107   &    1.20$\pm$0.22   &    0.30$\pm$0.02\\ \\
          &    &2    &    -33.83$\pm$0.21    &    24.95$\pm$0.22   &    -4.44$\pm$0.15   
          & 5.87$\pm$1.10   &  1891$\pm$64   &    3.91$\pm$0.13   &    0.35$\pm$0.00\\ \\ 
          &  1408&1    &    -14.70$\pm$0.40    &    11.78$\pm$0.34   &    -1.46$\pm$0.27   
          &    3.30$\pm$1.80   &    533$\pm$97   &    1.10$\pm$0.20   &    0.37$\pm$0.03\\ \\
          &  &2    &    -28.47$\pm$0.31    &    22.48$\pm$0.34   &    -2.99$\pm$0.23   
          &   5.58$\pm$1.80 &    1151$\pm$87   &    2.38$\pm$0.18   &    0.42$\pm$0.01\\ \\
\hline
\\
\end{tabular}

\tiny{$^a$Emission heights computed using the  exact formula (G05).}
\end{table*}

\subsection{Emission height of the cones}
We found the emission height of the cones based on the procedure that
is described in Sect.~\ref{subsec:comppeaks}.  In Col.~3 of
Table~\ref{tabcomp} we have given the cone numbers and the peak
locations of the conal components on the leading and trailing sides in
Cols.~4 and 5, respectively.  The conal components are believed to
arise from the nested conal emissions (Rankin 1983a, 1983b, 1990,
1993), which along with the central core emission make up the pulsar
emission beam.  In Col.~6 of Table~\ref{tabcomp}, we have given the
values of $\delta\phi'.$ In general it increases in magnitude from the
innermost cone to the outer cone.  The half-opening angle $\Gamma$
(see Eq.~(7) in G04) of the emission beam is given in Col.~7.  Using
the exact formalism for the A/R phase shift (see Eq.~(45) in G05), we
computed the emission heights and show them in Col.~8. Their percentage
values in $r_{\rm LC}$ are given in Col.~9.  In Col.~10, we 
give the normalized co-latitude $s/s_{\rm lof}$ of the foot field
lines on the polar cap, which are associated with the component
emissions. Due to the relativistic beaming and restrictions owing to
geometry, we find that the observer tends to receive the emissions
from open field lines, which are located in the foot locations ranging
from (approximately) 0.1 (Table 1) to 0.8 (Table 2) on the polar cap.

\subsection{PSR B1839+09} 
By fitting three Gaussians to the mean intensity profile of PSR
B1839+09 we have identified three sub-pulse components: a central
component and a pair of outer components flanking the central one.  By
invoking the picture of a nested cone structure we infer that the outer
pair of components corresponds to a conal emission. The inner component
might be due to emissions close to the magnetic axis or because of the
`grazing-cut' of the line-of-sight with an inner ring of emission. The
point of emission for the central component should fall in the
meridional plane M in the co-rotating frame of the pulsar in either of
the cases.  We  clearly identified the phase locations for the PPAIP
and the CP relative to the meridional plane M for PSR
B1839+09. The absolute emission heights estimated for the core and
conal component are given in Tables~\ref{tabcore} and~\ref{tabcomp},
respectively. The emission heights of the central (core) component and
cone are found to be $\sim 50$~km and $\sim 60$~km, respectively.

\subsection{PSR B1916+14}
By fitting Gaussians to the intensity profile, we could identify three
sub-pulse components: a central component and a pair of outer
components flanking the central one.  We  clearly identified the
phase locations of the PPAIP and the central peak, and found the phase of
meridional plane M. The absolute emission heights, estimated for the
central (core) and conal components, are given in Tables~\ref{tabcore}
\& \ref{tabcomp}. The emission heights of the central component and
the cone are estimated to be $\sim 100$~km and $\sim 300$~km,
respectively.
 
\subsection{PSR B2111+46} 
It is a well-studied pulsar.  By fitting Gaussians to its intensity
profiles at frequencies 610 MHz and 1408 MHz, we could identify five
sub-components: a central (core) component, a pair of inner components
and a pair of outer components.  One can guess at the presence of two
inner components hidden between the core and the outer components even
by visual inspection. These two hidden components were detected in the
previous work of GG03 in the 333~MHz single pulse data.  The absolute
emission heights estimated for the core and the cones are given in
Tables~\ref{tabcore} and~\ref{tabcomp}. The core emission height is
found to be $\sim 500$~km at 610 MHz and $\sim 80$~km at 1408 MHz.  It
has been reported in ML04 that the widths of the core tend to show a
significant frequency evolution in their chosen set of six pulsars,
and hence they argued that the core emission does not come from the
stellar surface. However, we need to consider more high quality data
sets at different frequencies to see the frequency evolution of the core
emission height, and whether it follows any radius--to--frequency
mapping.
                 
\section{Results and discussions}
Based on the A/R method, we estimated the absolute emission height of
the core as well as cones in three pulsars: PSRs B1839+09, B1916+14
and B2111+46. Though this method is based on the existing standard
models in literature, the combination of the A/R phase shift and the
delay-radius relation of BCW91 for estimating the core height is
novel.
 
The geometrical method, involving a comparison of the measured pulse
widths with geometrical predictions from dipolar models, is believed
to yield absolute emission heights. However, the estimation of
emission height, using the geometrical method, is based on the
assumption that the pulse edges originate from the last open field
lines of the polar cap.  In general, the edge of the on-pulse region
may not originate from the last open field line, and hence the
assigning of the edges of the intensity profile to the last open field
lines can be misleading.  For example, the range of magnetic
foot-colatitude for field-lines that are associated with components in
PSR B2111+46 are in the range from $S/S_{\rm lof}\sim 0.13$ to $0.5,$
whereas the last open field line is at $ 1.$ This means that the
boundary of the active region of emission can lie anywhere from $\approx
0.5$ to $1.$
 
According to DRH04, the A/R phase shift advances the centroid of the
intensity profile to an earlier phase by $\delta\phi'_c=2 r_{\rm
  lof}/r_{\rm LC},$ while the PPAIP is delayed to a later phase by
$\delta\phi'_{\rm PPAIP}\sim 2\,r_{\rm core}/r_{\rm LC},$ where
$r_{\rm lof}$ is the emission height from the last open
field-line and $r_{\rm core}$ is the emission height of the core.
Then $\Delta \phi'= 2 (r_{\rm lof}+ r_{\rm core})/ r_{\rm LC},$ and
the emission height $r= r_{\rm LC}\,\,\Delta \phi'/4 = (r_{\rm lof}+
r_{\rm core})/2 ,$ gives only an average of the emission height for
the core and the pulse edge, which is far from the true value.  This
emission height cannot represent any specific pulse sub-component of
the profile, and can be misleading in cases where $r_{\rm lof}$ and
$r_{\rm core}$ are significantly different. Further more, this will
introduce large systematic errors in the emission heights estimated
from geometrical methods, due to the aforesaid assumption of
identifying the last open field lines with the pulse edges.
   
Rankin (1983a) has argued that the pulsar emission cones are
quasi-axial, i.e., the conal components are not exactly axially
located with respect to the magnetic axis.  Mitra \& Deshpande (1999)
have suggested that the pulsar emission beams are nearly circular in
the aligned configuration ($\alpha\sim 0^\circ$) and change to
elliptical in the orthogonal configuration $(\alpha\sim90^\circ)$.
The majority of the pulsar observations indicate that the beam
geometry is likely to be nested cones, distributed in a nearly
non-coaxial fashion about the magnetic axis.  A likely case is that
the cones, which are coaxial in the co-rotating frame, will appear
non-coaxial in the laboratory frame because of the A/R phase shifts
(GG01; GG03).

In the works GG01 and GG03, the emission height of the core was
neglected by assuming that it is considerably smaller than that of the
components. However, we find that the emission height of the core is
quite significant and cannot be neglected in comparison to the
emission height of the components. We identify the meridional plane M
as being located at the mid point between the centroid of the
intensity profile and the PPAIP, owing to the A/R effects. By recognizing
this, we were able to estimate the absolute emission heights of both
the core and the conal components.

As mentioned before, we restricted the region of the fit of the BCW91
(relativistic RVM) curve to the section of the PPA data falling within
the FWHM of the core component for estimating the core emission
height, and the justification for doing so is given now. The
expression for the BCW91 was derived by assuming that the emission
altitude across the active region of the pulse profile is a
constant. Thus in a BCW91 fitting, a single $r$ value was taken to
characterize the emission height of the full region of the PPA
data. But later observational results (e.g., GG01, GG03) established
that the emission altitude corresponding to the subpulse components in
multi-component profiles spans over a large range of emission
heights. This elicits the fact that in multi-component profiles the
$r,$ found by fitting the BCW91 curve to the full PPA region of the
active profile, might give an emission altitude that can be
significantly different from those obtained from the A/R method for
the subpulse components.

The best-fit value of $r$ in the BCW91 model, which is the weighted
average of $r_{\rm i}$ that characterize the emission height at each
point of pulse phase, is given by Eq.~(\ref{eq_rbyrL2}) in
\ref{subsec:BCW}. Hence a single value of $r$ found from the BCW91 fit
cannot be closer to the true emission height corresponding to the core
or cone peak if $r_{\rm i}$ varies significantly within the region of
the fit.  For example, one can compare the emission altitudes in our
Tables \ref{tabcore} and \ref{tabcomp} with those given in Table~3 of
ML04 for PSR B2111+46. It can be surmised that a single value of $r$
cannot characterize the emission altitude across the entire active
region of a multi-component profile.

 We can think of two viable alternatives in this scenario: either (1)
adapt or modify the BCW91 formulation for a variable
emission altitude $r$ (Dyks 2008) or (2) fit the BCW91 curve for
regions of the PPA profile having a relatively constant value of $r.$
We prefer the latter alternative  to evade the modification of
the theory behind the BCW91 model.  Here we note that \emph{$\alpha$ and
  $\beta$ are not invoked as fit parameters}; instead we  used
their published values in Eq.~(\ref{eq_{BCW}}). This is expected to
further reduce the ambiguity of the fit results and to aid in
counteracting an obvious disadvantage in this `restricted' fit method,
i.e., having a reduced number of fitted PPA data points than a fit for
the `full range' of PPA data. It is remarkable that some of the fit
statistics (e.g., reduced $\chi^2$) given in Table~\ref{tabcore}
reveal that the present method of fitting is comparable (in a few
cases even better) to the existing ones in the literature (e.g.,
compare the $\chi^2$ value given in Table~\ref{tabcore} with Table~2
of ML04 for PSR B2111+46).  We  estimated the standardized
residuals (SR) and found the percentage of SR that falls within -2 and
+2 as given in Col.~(8) of Table~\ref{tabcore}. As it is known, a
good fit is expected to have a threshold 95 \% of the standardized
residuals to fall within -2 and +2.  Several previous works  found
that $\alpha$ and $\beta$ are highly covariant in PPA fits (e.g.,
Everette \& Weisberg 2001). But this covariance of $\alpha$ and
$\beta$ with the $r$ parameter was not mentioned by any of them. The fit
statistics do not reveal any significant covariance of $\alpha$ and
$\beta$ with $r.$ This gives us a further clue for finding the $r$
parameter without invoking a concurrent fit for $\alpha$ and $\beta$
(see \ref{app:Psi} for the fitting procedure).
                                    
Owing to the extreme difficulties encountered in determining $\alpha $
and $\beta$ through RVM fitting, a larger range of PPA data have
always been preferred for a better fit (e.g., Everette \& Weisberg
2001).  The justification for doing so is that $\alpha $ and $\beta$
must remain constant throughout the entire PPA profile. But in the
present scenario, as described earlier, the selection of a large range
of PPA data for fitting does not always translate into a better
estimation of $r$ because of the variation of the emission height with
pulse phase. So, owing to all of the above said reasons we restrict
the fit of the BCW91 curve to the PPA profile, falling around the core
component, which is expected to yield an emission altitude
characterizing the core height.

A section of inner cones often lapses over the core as is seen in the
Gaussian fits (panel (a) of Figs.~\ref{freq_1839}--\ref{freq_1408}) of
the total intensity profiles. The inner cones may contribute to the
core polarization near the edges of pulse phase of the FWHM region
that we bracketed.  Hence the PPA corresponding to the bracketed
region will be `contaminated ' by the adjacent conals, and this has to
be accounted for.  We estimated and accounted for the error induced
because of this effect in the estimation of the core emission heights
(see \ref{app:Psi1} and \ref{subsec:BCW}).
 
The possibility that the A/R phase shift may be reduced by the
rotational distortion of the magnetic field line due to a sweep-back
of the vacuum dipole magnetic field lines has to be considered.  The
sweep-back of dipole magnetic field lines was first treated in detail
by Shitov (1983). Further, Dyks \& Harding (2004)  investigated
the rotational distortion of pulsar magnetic field by making the
approximation of a vacuum magnetosphere. For $\phi'=30^\circ,$
$\beta=-1.6^\circ$ and $\alpha = 14^\circ$ we computed the phase shift
$\delta\phi'_{\rm mfsb}$ due to the magnetic field sweep-back (Dyks \&
Harding 2004; also see Eq. (49) in G05).  It is found to be
$<0.0001$~rad for $r/r_{\rm LC}\leq 0.06, $ which is much smaller than
the aberration, retardation and polar cap current phase shifts in
PSR~B2111+46.  Hence we neglect the magnetic field sweep-back effect.

The field-aligned polar-cap current does not introduce any significant
phase shift into the phase of the PPAIP. But it introduces a positive
offset into the PPA, though it roughly cancels due to the negative
offset by aberration (Hibschman \& Arons 2001). The phase shift of
pulse components due to the polar cap current was estimated recently by
G05, and found to be quite small compared to the A/R phase shift.

\section{Summary} 
We analyzed the mean profiles of PSRs B1839+09 and B1916+14 at 1418
MHz, and those of B2111+46 at 610 MHz and 1408 MHz.  The phase of the
peak of central component (core) and that of the polarization position
angle inflection point are symmetrically shifted in the opposite
directions with respect to the meridional plane due to A/R effects.
By recognizing this, {\it we were able to locate the phase of the
  meridional plane and to estimate the absolute emission altitudes of
  the core and the conal components relative to the center of the
  neutron star.}  We used the exact expression for the phase shift
given recently by G05. In all the cases we found that the core
emission occurs at a relatively lower altitudes than the conal
emissions.  It is also interesting to note that the core emission at
different frequencies in PSR B2111+46 falls in a range of altitude of
80~km at 1408 MHz to about 500~km at 610 MHz. It is clear that the low
frequency emission comes from a higher height than that at high
frequency. However, to confirm whether the core emission heights also
obey any radius-to-frequency mapping demands the recursive analysis
with high quality multi-frequency data. We plan to employ the methods
described in this paper for the study of a few other pulsars with high
quality data.
                            
\begin{acknowledgements}
  We thank Joel Weisberg for providing the data of PSRs B1839+09 and
  B1916+14 at 1418~MHz.  We used the data available on EPN archive
  maintained by MPIfR, Bonn, and thank all the observers who have made
  their data available on the EPN data base.  We thank Yashwant Gupta
  for the helpful comments.
\end{acknowledgements}

\clearpage         
\newpage            
\Online     
\renewcommand{\theequation}{A-\arabic{equation}}
\setcounter{equation}{0}   
 \appendix    
\renewcommand\thesection{Appendix \Alph{section}} 
\section{Estimation of $1\,\sigma_{\Psi}$}
\label{app:Psi}    
We apply the standard methods of statistics for fitting the 
observed data with the model. For fitting 
the PPA data with the BCW91 curve we define the reduced  $ \chi^2$ as  
\begin{equation}\label{eq_chisq}
 \chi^2= \sum_{i}^{\rm N} \left[\frac{\psi_{\rm POL}(\phi'_{\rm i})-
\psi_{\rm   BCW}(\phi'_{\rm i})}{\sigma_{\psi}(\phi'_{\rm i})} \right]^2 ~,
\end{equation} 
where $\Psi_{\rm POL}$ is the observed polarization angle and
$\Psi_{\rm BCW}$ is the model (BCW91) value at the discrete 
rotation phase $\phi_{\rm i}'$, and $N$ is the number of data
points. Minimization of  $ \chi^2$ gives the best fit. We use
Mathematica version~7 for the fitting and other calculations that
ensue therewith.
 
For profile regions with a very high value of $L/\sigma_{\rm I},$ i.e., for $
L/\sigma_{\rm I} \gg 10, $ the $\sigma_{\psi} $ is taken as
\begin{equation} 
   \sigma_{\psi}=\frac{\sqrt{U^2\,\sigma_{\rm Q}^2+Q^2\,\sigma_{\rm U}^2}}{2 L}~,
\end{equation}
where $\sigma_{\rm Q}$ and $\sigma_{\rm U}$ are the RMS values of the
off-pulse $Q$ and $U$ stokes parameters, respectively. The parameter
$L=\sqrt{Q^2+U^2}$ represents the linear polarization and $\sigma_{\rm
  I}$ is the standard deviation of  total intensity in the
off-pulse region.  Though the measurements of $U$ and $Q$ around their
true values are normally distributed, the PPA measurements are not
similarly distributed at the intermediate values of $L/\sigma_{\rm I}.$
Hence the form of $\sigma_{\psi} $ as given above is not appropriate
in this regime.  The appropriate distribution function that
characterizes the $\psi$ distribution is discussed in detail by
Naghizadeh-Khouei \& Clarke (1993).  The application of their scheme is
discussed by Everette \& Weisberg (2001); we follow their method for
finding the $\sigma_{\psi}$ for the PPA data points.  The probability
distribution function of the measured position angle $\psi$ around a
true value $\psi_{\rm true}$ is given by Naghizadeh-Khouei \& Clarke
(1993) as 
 \begin{equation}\label{eq_prob}  
 G(\psi;\, \psi_0,\, P_0) = \frac{1}{\sqrt{\pi}}\left(\frac{1}{\sqrt{\pi}} +
 \eta_0 \, e^{\eta_0^2}(1 + erf(\eta_0))\right) e^{-\frac{P_0^2}{2}}~,
\end{equation}   
 where $\eta_0=(P_0/\sqrt{2}) \cos[2(\psi-\psi_{\rm true})],$ $P_0=
 L_{\rm true}/\sigma_{\rm I}, $ $erf$ is the error function and $
 L_{\rm true}$ is the unbiased linear polarization found from the
 measured linear polarization.  Hence the $1\sigma_{\psi}$ confidence
 level in the PPA is found out by adjusting the limits of integration
 and fixing $\psi_{\rm true}=0,$
 \begin{equation}
 \int^{1\sigma_{\psi}}_{-1\sigma_{\psi}} G(\psi;P_0)\, d\psi=62.86\%. 
 \end{equation}
   We set up a table of the integration values of the integral against
   a series of discrete $P_0$ values and the $ 1 \sigma_{\psi} $
   levels at integral value of 0.6286 are found by interpolation.
\renewcommand\thesection{Appendix \Alph{section}} 
\section{Polarization angle $\Psi$: Contribution from the
 adjacent component to the core}
\label{app:Psi1}    

The polarization angle $\Psi$ is defined as
\begin{equation} \label{eq_chi}
\Psi=\frac{1}{2}\tan^{-1}\left(\frac{U}{Q}\right)\,.
\end{equation}
Defining
\begin{eqnarray} 
  U &= & U_0+U_1 \label{eq_U}~,\\ 
Q &= & Q_0+Q_1  \label{eq_Q}~,\\
L&= &\sqrt{U^2+Q^2}\label{eq_L}~, \\
 L_0&= &\sqrt{U_0^2+Q_0^2}~,\label{eq_L}
\end{eqnarray}
where $U_1$ and $Q_1$ are taken as sufficiently small, $\Psi$ can be a
series expanded up to the first order in $U_1$ and $Q_1$ as
\begin{eqnarray}
\Psi &\simeq& \Psi_0+\Delta\Psi \label{eq_chi_all}~,\\  
\Psi_0 &=& \frac{1}{2} \tan^{-1}\left(\frac{U_0}{Q_0}\right)\label{eq_chi_0}~,\\
\Delta\Psi &=&\frac{1}{2} \frac{Q_0U_1-U_0Q_1}{U_0^2+Q_0^2}\label{eq_chi_1}~.
\end{eqnarray}  
    
The expression for $\Delta\Psi $ may be approximated as 
\begin{eqnarray} 
 \Delta\Psi &\simeq& \frac{1}{2} \frac{Q\,U_1-U\,Q_1}{L^2}\label{eq_chi_1}~,
\end{eqnarray}  
provided $U_1$ and $Q_1$ are sufficiently small, so that $L_0\simeq L$.  
 
  We use the above expressions for estimating the contribution to
  $\Psi$ from the adjacent conal components. The suffix `1' indicates
  the $U$ and $Q$ contribution solely from the inner cone, while the
  suffix`0' indicates the pure core contribution for the same. We
  employ approximations to find the value of $U_1$ and $Q_1.$
   
 Since the total intensity of an inner conal component can be fitted
 with a Gaussian, we make an assumption that the $U_1(\phi')$ and
 $Q_1(\phi')$ also follow a Gaussian shape within the adjacent conal
 component. Thus
      \begin{equation} 
U_1({ \phi'}) =  U_{1({\rm max})} \exp\left(-\frac{(\phi'_{\rm max}
 - { \phi'})^2}{\sigma^2_{\phi}}\right)\label{eq_U1}
\end{equation}  and 
 \begin{equation}  
Q_1({ \phi'}) =  Q_{1({\rm max})} \exp\left(-\frac{(\phi'_{\rm max}
 - {\phi'})^2}{\sigma^2_{\phi}}\right)\label{eq_Q1}~,  
\end{equation}        
where $\sigma_{\phi} $ is the width and $\phi'_{\rm max}$ is the peak
phase of the inner conal component (same as the corresponding Gaussian
for the total intensity), while $ U_1({\rm max})$ and $Q_1({\rm max})$
are the peak values of $U_1$ and $Q_1,$ respectively. The total
intensity of the inner cone at phase $\phi'_{\rm int},$ where the
Gaussians corresponding to the core and outer cone intersect, will
have the minimum. Hence the values of $U_1(\phi'_{\rm int}) $ and
$Q_1(\phi'_{\rm int})$ may represent approximately the true values of
$U_1$ and $Q_1$ for the inner-cone at $\phi'_{\rm int}.$ By using
Eqs.~(\ref{eq_U1}) and  (\ref{eq_Q1}) we can write
 \begin{eqnarray}\label{eq_peaks_UQ}  
U_{1({\rm max})} &=&U_1(\phi'_{\rm int})\,\,\exp\left(
    \frac{(\phi'_{\rm max}-\phi'_{\rm int})^2 }{\sigma^2_{\phi}}
    \right) \label{eq_peaks_U1}  ~, \\
Q_{1({\rm max})} &=& Q_1(\phi'_{\rm int})\,\,\exp\left(
   \frac{(\phi'_{\rm max}-\phi'_{\rm int})^2 }{\sigma^2_{\phi}}
    \right)~.\label{eq_peaks_Q1}  
\end{eqnarray}    
Using the Eqs.~(\ref{eq_peaks_U1}) and (\ref{eq_peaks_Q1}), we can
find $U_1({ \phi'})$ and $Q_1({ \phi'}),$ and hence the $\Delta\Psi({
  \phi'})$ is estimated within the bracketed region of PPA.

{\bf Note:} The PSR B1839+09 has practically no contribution of
polarization from the adjacent conals to the bracketed core
region. Hence this analysis is not performed for it. The $U_1({\rm
  max}) $ and $Q_1({\rm max}) $ for PSR B1916+14 are the values at the
peak phases of the adjacent cones and are directly found from the
profile data. So, the Eqs.~(\ref{eq_peaks_U1}) and
(\ref{eq_peaks_Q1}) are not applicable for it. Hence the above-said
analysis is done for the profiles of PSR B2111+46 only.

\renewcommand\thesection{Appendix \Alph{section}} 
\section{Determination of the $r$ parameter: Correcting for adjacent 
          conal contribution}
\label{subsec:BCW}  
Here, we explain the scheme of the determination of $r$ and
$\delta\phi'_{\rm PPAIP} $ using the BCW91 method with predetermined
values of $\alpha$ and $\beta$.  Consider the expression for the
polarization angle given in BCW91:
 \begin{equation}\label{eq_{BCW}}    
 \psi_{\rm BCW}(\phi_{\rm i})= \tan^{-1}  
   \left[\frac{\sin\alpha \, \sin\phi_{\rm i} -3 (r/r_{\rm LC})
       \sin\zeta}{\sin\beta+\sin\alpha\cos\zeta(1-\cos\phi_{\rm i})}\right]~.
 \end{equation}  
  Here define   $ \chi^2$   as 
   \begin{equation}\label{eq_chisq_new}  
 \chi^2= \sum_{i}^{\rm N} \left[\frac{\tan[\, \psi_{\rm POL}(\phi'_{\rm i})]-
\tan[ \, \psi_{\rm   BCW}(\phi'_{\rm i})]}{\sigma_{\tan\psi}(\phi'_{\rm i})} \right]^2 ~,
\end{equation}  
where $\sigma_{\tan\psi}(\phi'_{\rm i})= \tan[\sigma_{\psi}(\phi'_{\rm
    i})],$ $\sigma_{\psi}(\phi'_{\rm i}) $ is the $1\sigma$ error
determined by using the Naghizadeh-Khouei \& Clarke (1993) method
(\ref{app:Psi}). The definition of $\chi^2$ as given in
Eq.~(\ref{eq_chisq_new}) is slightly different from that used in the
fitting, and this is meant exclusively for an explanatory
purpose. Using the $\chi^2$ of this form makes it easy to derive the
following expressions for the $r$ parameter from the BCW91 model. The
form of $\chi^2$ that is used in the actual fitting is given in
\ref{app:Psi} (Eq.~\ref{eq_chisq}), and the method of correction for
the adjacent conal contribution is briefly given at the end of this
section.

By minimizing the $ \chi^2$ with respect to  the parameter $r,$ we can write
  \begin{eqnarray}   
 \frac{d\,\chi^2}{d\,r} &=& 2 \sum_{i}^{\rm N} \left(\frac{\tan(\, 
     \psi_{\rm POL}(\phi'_{\rm i}))- \tan(\, \psi_{\rm 
       BCW}(\phi'_{\rm i}))}{\sigma_{\tan\psi}(\phi'_{\rm i})} \right)\nonumber \\
&&\times  \frac{d\,}{dr} \tan( \, \psi_{\rm BCW}(\phi'_{\rm i})) =0~.\nonumber 
\end{eqnarray}  
 By substituting for $\tan(2 \, \psi_{\rm BCW}(\phi'_{\rm i})), $ we obtain
 \begin{equation}\label{eq_rbyrL}   
 r = \frac{r_{\rm LC}}{3}  \sum_{i}^{\rm N} \frac{ F(\phi'_{\rm i}) F_1(\phi'_{\rm i})
 }{\sigma^2_{\tan\psi}(\phi'_{\rm i})} 
 \Big{/}  \sum_{i}^{\rm N}
   \frac{\sin\zeta\,\, F^2(\phi'_{\rm i})  }{\sigma^2_{\tan\psi}(\phi'_{\rm i})} ~, 
\end{equation}    
where       
\begin{equation}  
 F_1(\phi'_{\rm i})  =   \frac{ \sin\alpha \,
   \sin\phi'_{\rm i}}{\sin\beta+\sin\alpha\cos\zeta(1-\cos\phi'_{\rm
     i})} -\tan(\psi_{\rm POL}(\phi'_{\rm i}))
\end{equation}
and 
\begin{equation} 
F(\phi'_{\rm i}) =   
\frac{1}{\sin\beta+\sin\alpha\cos\zeta(1-\cos\phi'_{\rm i})}~. 
\end{equation}   
   
If $r_{\rm i}$ is a  value so that $\psi_{\rm POL}(\phi_{\rm 
  i})=\psi_{\rm BCW}(\phi_{\rm i}), $ 
$(r_{\rm i}$ may characterize the true emission height at the phase  $\phi_{\rm i}),$
then by taking the analogy from Eq.~(\ref{eq_{BCW}}) we can equate
 \begin{equation}\label{eq_rbyrLTrue} 
  F_1(\phi'_{\rm i}) =   \frac{3\,r_{\rm i}}{r_{\rm LC}}\,\,\, 
   \frac{\sin\zeta}{\sin\beta+\sin\alpha\cos\zeta(1-\cos\phi'_{\rm
       i})}~. 
\end{equation}      
 Then by substituting Eq.~(\ref{eq_rbyrLTrue}) into Eq.~(\ref{eq_rbyrL}), we get
 \begin{equation}\label{eq_rbyrL1}           
 r =  \sum_{i}^{\rm N} r_{\rm i}  
     \frac{F^2(\phi'_{\rm i})}{\sigma^2_{\tan\psi}(\phi'_{\rm i})} {\Big{/}} 
     \sum_{i}^{\rm N}   
      \frac{ F^2(\phi'_{\rm i})}{\sigma^2_{\tan\psi}(\phi'_{\rm i})}.
\end{equation}    
  Within the bracketed region the small angle approximation for $\phi_{\rm i}'$
can be used so that  we can re-write  expression (\ref{eq_rbyrL1}) as 
 \begin{equation}\label{eq_rbyrL2}            
 r \simeq    \sum_{i}^{\rm N} r_{\rm i}   
     \frac{1}{\sigma^2_{\tan\psi}(\phi'_{\rm i})} {\Big{/}}  
     \sum_{i}^{\rm N}   
      \frac{1}{\sigma^2_{\tan\psi}( \phi'_{\rm i})} ~, 
\end{equation}     
or, by    using $\delta\phi'_{\rm PPAIP}= 2 r/r_{\rm LC}$ and 
 $\delta\phi'_{\rm i\,(shift)}= 2 r_{\rm i}/r_{\rm LC}$  we can write 
 \begin{equation}\label{eq_rbyrL3}           
\delta\phi'_{\rm PPAIP}  \simeq    \sum_{i}^{\rm N}\delta\phi'_{\rm i\,(shift)}
     \frac{1}{\sigma^2_{\tan\psi}(\phi'_{\rm i})} {\Big{/}} 
     \sum_{i}^{\rm N}      
      \frac{1}{\sigma^2_{\tan\psi}(\phi'_{\rm i})} ~.   
\end{equation}   
These expressions (\ref{eq_rbyrL2}) and (\ref{eq_rbyrL3}) show that the
BCW fit of the PPA data is a weighted average of the emission heights
$r_{\rm i}$ of each data point.  If some of the data points are
corrupted by the adjacent conals, then the true linear polarization due to the
core alone should be less than the observed. Hence it implies that the
true PPA value corresponding to the pure core contribution should be
slightly different from the observed PPA, which has a small mixture of
contribution from adjacent conal component.  An approximate method to
estimate this error in the PPA ($\Delta\Psi$) due to the `contamination'
of the adjacent core is explained in a previous section. Hence  the new
weight factors are to  be defined as $$\sigma_{\rm TOTAL}(\phi'_{\rm i})
=\sqrt{\sigma_{\tan\psi}^2(\phi'_{\rm i}) +\tan^2[\Delta\Psi_{\rm
      cone}(\phi'_{\rm i})]}~,$$ 
which takes into account the `contamination' of the core due to the
adjacent cones.  Thus we find
     \begin{equation}\label{eq_rbyrL4}            
 r^{\rm pure} \simeq \sum_{i}^{\rm N} r_{\rm i} \frac{1}{\sigma^2_{\rm
     TOTAL}(\phi'_{\rm i})} {\Big{/}} \sum_{i}^{\rm N}
 \frac{1}{\sigma^2_{\rm TOTAL}( \phi'_{\rm i})}
\end{equation}  
 and   
   \begin{equation}\label{eq_rbyrL5}           
 \delta\phi'^{\rm pure}_{\rm PPAIP} \simeq \sum_{i}^{\rm N} \delta\phi'_{\rm shift}
 \frac{1}{\sigma^2_{\rm TOTAL}(\phi'_{\rm i})} {\Big{/}} \sum_{i}^{\rm
   N} \frac{1}{\sigma^2_{\rm TOTAL}( \phi'_{\rm i})} 
\end{equation}
using the expressions (\ref{eq_rbyrL2}) and (\ref{eq_rbyrL3}).  In
principle, $ r^{\rm pure}$ and $\delta\phi'^{\rm pure}_{\rm PPAIP} $
characterize the values of emission height and the corresponding phase
shift of the PPAIP found for the core region after correcting for the
conal contamination. The error induced in locating the PPAIP due to
the adjacent conal contamination can be given as
$\Delta\delta\phi'_{\rm PPAIP}= |\delta\phi'^{\rm pure}_{\rm
  PPAIP}-\delta\phi'_{\rm PPAIP}|. $
 Hence the  improved value of the phase location  of  PPAIP can be expressed  as 
 \begin{equation}\label{eq_PURE_PPAIP}
\delta\phi'_{\rm PPAIP}\pm \sqrt{{\rm Error}(\delta\phi'_{\rm     
      PPAIP})_{\rm fit}~^2 + {\Delta\delta\phi'~^2_{\rm PPAIP}} }~~,  
  \end{equation}
   where ${\rm  Error}(\delta\phi'_{\rm PPAIP})_{\rm fit} $ is the
   $1\sigma$ confidence interval for the $\delta\phi'_{\rm PPAIP} $
   obtained from the BCW91 fit without invoking the correction for conal 
 contamination. 

In the actual fitting procedure we  used the form of $\chi^2$ as
given in \ref{app:Psi}.  We  estimated $|\Delta\delta\phi'_{\rm
  PPAIP}|$ for the profiles, and the values of $\delta\phi'_{\rm
  PPAIP} $ (Table~1) were attributed with an error factor as
given by Eq.~(\ref{eq_PURE_PPAIP}).  A set of weights are found in the
form of $\sigma_{\rm TOTAL} =\sqrt{\sigma_{\psi}^2 +\Delta\Psi_{\rm
    cone}^2}.$ The BCW91 model is fitted within the bracketed region
after (1) weighting the data with weights $\sigma_{\psi} $ and (2) again
by weighting the data with weights $\sigma_{\rm TOTAL} .$ The case
(1) will yield the phase shift for the PPAIP, while the case (2) should
yield the phase shift for the PPAIP with reduced weights to the PPA
points where the conal contributions are present.  The difference in
the phase shifts found by case (1) and (2) should characterize the
extra increment (decrement) in core emission height due to the adjacent
conal contributions. The square of this phase shift difference is
added with the squared error of the PPAIP phase shift, and the square root
of this sum will give the improved error factor for the core emission
height.  This improved error factor will take into account the
error induced in the estimation of the PPAIP in the bracketed region
due to the adjacent conal contribution.


\begin{thebibliography}{} 
\bibitem[]{} Blaskiewicz, M., Coders, J. M., \& Wasserman, I. 1991,
  \apj, 370, 643, (BCW91)
\bibitem[]{} Cordes, J. M. 1978, \apj, 222, 1006
\bibitem[]{} Dyks, J. 2008,\mnras  391, 859
\bibitem[]{} Dyks, J., Rudak, B., \& Harding, A. K. 2004, ApJ, 607,
  939, (DRH04)
\bibitem[]{} Dyks, J., \& Harding, A. K. 2004, \apj, 614, 869
\bibitem[]{} Everette, J.E., \& Weisberg, J.M., 2001, \apj, 341, 357
\bibitem[]{} Gangadhara, R. T. 2004, \apj, 609, 335, (G04)
\bibitem[]{} Gangadhara, R. T. 2005, \apj, 628, 930, (G05)
\bibitem[]{} Gangadhara, R. T. \& Gupta, Y. 2001, ApJ, 555, 31, (GG01)
\bibitem[]{} Gil, J. A. \& Kijak, J. 1993, \aap, 273, 563
\bibitem[]{} Gupta, Y.  \& Gangadhara, R. T.  2003, \apj, 584, 41,
  (GG03)
\bibitem[Hibschman \& Arons(2001)]{HA01}Hibschman, J. A., \& Arons,
  J. 2001, ApJ, 546, 382
\bibitem[]{} Hoensbroech, von A. \& Xilouris, K. M.  1997, \aap, 324,
  981
\bibitem[Johnston (2006)]{J2006} Johnston, S. \& Weisberg, J. M. 2006,
  MNRAS, 368, 1856
\bibitem[]{} Kijak, J.  \& Gil, J. 2003, \aap, 397, 969
\bibitem[]{} Kramer, M., Wielebinski, R., Jessner, A., Gil, J. A., \&
  Seiradakis, J. H.  1994, A\&AS, 107, 515  
\bibitem[]{} Krzeszowski, K., et al. 2009, MNRAS, 393, 1617
\bibitem[]{} Lyne, A. G. \& Manchester, R. N. 1988, \mnras, 234, 477
\bibitem[]{} Mitra, D.  \& Deshpande, A. A.  1999, \aap, 346, 906 
\bibitem[Mitra \& Li (2004)]{ML04} Mitra, D. \& Li, X. H. 2004, \aap,
  421, 215, (ML04)
\bibitem[]{}Naghizadeh-Khouei, J., \& Clarke, D. 1993, \aap, 274, 968
\bibitem[Rankin (1983a)]{R83a} Rankin, J. M. 1983a, \apj, 274, 333
\bibitem[Rankin (1983b)]{R83b} Rankin, J. M. 1983b, \apj, 274, 359
\bibitem[Rankin (1990)]{R90d} Rankin, J. M. 1990, \apj,  352, 247
\bibitem[Rankin (1993)]{R93d} Rankin, J. M. 1993, \apjs, 85, 145
\bibitem[Ruderman \& Sutherland(1975)]{RS75}Ruderman, M. A., \&
  Sutherland, P. G. 1975, ApJ, 196, 51
\bibitem[Shitov (1983)]{shi} Shitov, Yu. P. 1983, Sov. Astron. 27, 314
\end{thebibliography}
\end{document}